# Impact of surface collisions on enhancement and quenching of the luminescence near the metal nanoparticles

Jacob B. Khurgin,[1] and Greg Sun,[2,*]

[1]Department of Electrical & Computer Engineering, Johns Hopkins University, Baltimore, MD 21218, USA
[2]Department of Physics, University of Massachusetts at Boston, Boston, MA 02125, USA
[*]greg.sun@umb.edu

**Abstract:** The fact that surface-induced damping rate of surface plasmon polaritons (SPPs) in metal nanoparticles increases with the decrease of particle size is well known. We show that this rate also increases with the degree of the mode confinement, hence damping of the higher order nonradiative SPP modes in spherical particles is greatly enhanced relative to damping of the fundamental (dipole) SPP mode. Since higher order modes are the ones responsible for quenching of luminescence in the vicinity of metal surfaces, the degree of quenching increases resulting in a substantial decrease in the amount of attainable enhancement of the luminescence.

**References and links**

1. M. Stockman, "Nanoplasmonics: past, present, and glimpse into future," Opt. Express **19**, 22029-22106 (2011)
2. S. A. Maier, Plasmonics: Fundamentals and Applications (Springer, 2007).
3. W. L. Barnes, A. Dereux, and T. W. Ebbesen "Surface plasmon subwavelength optics", Nature **424**, 824 (2003)
4. K. Okamoto, I. Niki, and A. Scherer, Appl. Phys. Lett. **87**, 071102 (2005)
5. S. Kühn, U. Håkanson, L. Rogobete, and V. Sandoghdar, "Enhancement of single-molecule fluorescence using a gold nanoparticle as an optical nanoantenna," Phys. Rev. Lett. 97, 017402 (2006)
6. L. Novotny, "Single molecule fluorescence in inhomogeneous environments," Appl. Phys. Lett. **69**, 3806 (1996)
7. S. Kühn, G. Mori, M. Agio, and V. Sandoghdar, "Modification of single molecule fluorescence close to a nanostructure: radiation pattern, spontaneous emission and quenching," Mol. Phys. **106**, 893 (2008)
8. P. Bharadwaj and L. Novotny, "Spectral dependence of single molecule fluorescence enhancement," Opt. Express **15**, 14266-14274 (2007)
9. R. Bardhan, N. K. Grady, J. R. Cole, A. Joshi, and N. J. Halas, "Fluorescence enhancement by Au nanostructures: nanoshells and nanorods," ACS Nano **3**, 744-752 (2009)
10. S. Nie and S. R. Emory, "Probing single molecules and single nanoparticles by surface-enhanced Raman scattering," Science **275**, 1102-1106 (1997)
11. A. M. Michaels, M. Nirmal, and L. E. Brus, "Surface enhanced Raman spectroscopy of individual Rhodamine 6G molecules on large Ag nanocrystals," J. Am. Chem. Soc. **121**, 9932-9939 (1999)
12. Z. Wang, S. Pan, T. D. Krauss, H. Dui, and L. J. Rothberg, "The structural basis for giant enhancement enabling single-molecule Raman scattering," Proc. Natl. Acad. Sci. U.S.A. **100**, 8638-8643 (2003)
13. E. C. Le Ru, and P. G. Etchegoin, Principles of Surface Enhanced Raman Spectroscopy and Related Plasmonic Effects, (Elsevier, Amsterdam, 2009)
14. M. Bakker, V. P. Drachev, Z. Liu, H.-K. Yuan, R. H. Pedersen, A. Boltasseva, J. Chen, J. Irudayaraj, A. V. Kildishev, and V. M. Shalaev, "Nanoantenna array-induced fluorescence enhancement and reduced lifetimes," New J. Phys. **10**, 125022-1-16 (2008)
15. R. Carminati, J.-J. Greffet, C. Henkel, and J. M. Vigoureux, "Radiative and non-radiative decay of a single molecule close to a metallic nanoparticle," Opt. Commun **261**, 368-375 (2006)
16. G. Sun, J. B. Khurgin, and R. A. Soref, "Practical enhancement of photoluminescence by metal nanoparticles," Appl. Phys. Lett. **94**, 101103 (2009)
17. E. Dulkeith, A. C. Morteani, T. Niedereichholz, T. A. Klar, and J. Feldmann, "Fluorescence quenching of dye molecules near gold nanoparticles: radiative and nonradiative effects," Phys. Rev. Lett. **89**, 203002 (2002)
18. G. Sun, J. B. Khurgin, and C. C. Yang, "Impact of high-order surface Plasmon modes of metal nanoparticles on enhancement of optical emission," Appl. Phys. Lett. **95**, 171103 (2009)

19. G. Sun, J. B. Khurgin, and D. P. Tsai, "Comparative analysis of photoluminescence and Raman enhancement by metal nanoparticles", Opt. Lett. **37**, 1583-1585 (2012)
20. G. Sun, J. B. Khurgin, "Origin of giant difference between fluorescence, resonance, and nonresonance Raman scattering enhancement by surface plasmons", Phys. Rev. A **85**, 063410 (2012)
21. J. B. Khurgin, "How to deal with the loss in plasmonics and metamaterials", Nature Nanotechnology **10**, 2-6, (2015)
22. U. Kreibig and M. Vollmer, Optical Properties of Metal Clusters (Springer-Verlag, Berlin, 1995).
23. A.V. Uskov, I. E. Protsenko, N. A. Mortensen, and E. P. O'Reilly, "Broadening of plasmonic resonance due to electron collisions with nanoparticle boundary: a quantum mechanical consideration," Plasmonics **9**, 185–192 (2014);
24. C. Yannouleas, R.A. Broglia "Landau damping and wall dissipation in large metal clusters", Annals of Physics **217**, I 105–141 (1992)
25. R.A. Molina, D. Weinmann, R.A. Jalabert, "Oscillatory size dependence of the surface plasmon linewidth in metallic nanoparticles", Phys. Rev. B, **65**, 155427 (2002)
26. Z. Yan, S. Gao, "Landau damping and lifetime oscillation of surface plasmons in metallic thin films studied in a jellium slab model", Surface Science **602**, 460-464 (2008)
27. F. J. Garcıa de Abajo, "Nonlocal effects in the plasmons of strongly interacting nanoparticles, dimers, and waveguides," J. Phys. Chem. C **112**, 17983–17987 (2008).
28. N. A. Mortensen, S. Raza, M. Wubs, T. Søndergaard, and S. I. Bozhevolnyi, "A generalized nonlocal optical response theory for plasmonic nanostructures," Nat. Commun. **5** (2014).
29. G. D. Mahan, Many-particle physics (Kluwer Academics, New York, 2000)
30. P. B. Johnson and R. W. Christy, "Optical Constants of Noble Metals" Phys. Rev. B 6, 4370 (1972)
31. F. Tam, G. P. Goodrich, B.R. Johnson, N. J. Halas, "Plasmonic Enhancement of Molecular Fluorescence", Nano Lett. **7**, 496-501, (2007)
32. O. L. Muskens, V. Giannini, J. A. Sánchez-Gil, and J. Gómez Rivas, "Optical scattering resonances of single and coupled dimer plasmonic nanoantennas," Opt. Express **15**, 17736-17746 (2007)

**1. Introduction**

It is well known that emission properties of nanometer-scale objects, such as atoms, molecules, or quantum dots placed near metal surfaces get modified. This modification stems from interaction between the emitting object and collective oscillations of free-electrons in the metal called surface plasmon polaritons (SPP's) [1,2]. SPPs are characterized by strong electric fields localized near metal surfaces, and if the dimensions of the metal nanoparticles are sub-wavelength, so is the spatial extent of the localized field [3]. Enhancements of radiative decay rates [4], fluorescence [5-9], Raman scattering [10-12] and others near metal nanoparticles have been demonstrated by numerous groups. Spectacular enhancement of Raman scattering by metal nanoparticles with far more orders of magnitude [10-13] has since made surface enhanced Raman scattering (SERS) a practical sensing technology, but when it comes to the luminescence enhancement the progress has been less impressive [14]. Not only the emission efficiency suffers from non-radiative decay (damping) of SPP modes [15,16], but the situation is exacerbated by the process of life-time quenching [17] in which the energy of the emitting object ("emitter") gets coupled into the higher-order so-called "dark" SPP modes that do not radiate and hence dissipates inside the metal [18].

The modes are "dark" because the electric field in them undulates (changes sign) on much shorter scales than the wavelength of radiation, hence they are poorly coupled to the radiative modes. This is the case for the large wavevector propagating SPP modes on the metal-dielectric interface, or higher order localized SPP modes in metal nanoparticles, which are the subject of this work. Previously in [18-20] we have developed the theory of luminescence quenching due to excitation of higher order modes. We have shown that as emitting object approaches the metal surface, the coupling into higher modes gets enhanced relative to the coupling into the radiative lowest order (dipole) mode. As larger fraction of the energy gets coupled from the emitter into the dark modes, the luminescence gets quenched. Since the resonant frequencies of higher order modes are blue-shifted relative to the dipole mode, the coupling (and hence quenching) critically depends on the detuning and linewidth of SPP resonances.

In the first-order approximation SPP mode can be treated as a harmonic oscillator whose Lorentzian linewidth is equal to the damping rate $\gamma$ that has two constituents – radiative decay rate $\gamma_{rad}$ that depends on geometry (and is vanishingly small for all non-radiative higher-order modes), and the non-radiative rate $\gamma_{met}$ specific to the metal, the same rate that enters the Drude expression for the dielectric constant of the metal $\varepsilon_{met} = 1 - \omega_p^2/(\omega^2 + i\omega\gamma_{met})$. Using the language of classical physics, this "bulk" damping rate represents ohmic losses in the metal. Using the "quantum" language, one can describe it as a combination of processes of an SPP (or a photon) annihilation and creation of one and more electron-hole pairs in the metal. Since both energy and momentum must be conserved in this process, a mechanism that compensates for the momentum mismatch must be involved. This can be a scattering by phonons, defects, and impurities, or electron-electron scattering [21] and all the SPP modes experience the same "bulk" non-radiative damping $\gamma_{met}$. Yet as early as 1980's [22] it was demonstrated that as metallic objects become smaller there appears a new size-dependent contribution to the non-radiative damping which, classically speaking, can be associated with the electron scattering on the surface of nanoparticle. Using this phenomenological description Kreibig and Vollmer [22] introduced the "surface collision scattering rate" $\gamma_s \sim v_F/D$, where $D$ is a characteristic dimension of nanoparticle and $v_F$ is a Fermi velocity. On the quantum level this "surface collision damping" [23] is nothing but the Landau damping [24-26], a process in which large wave-vector components of the electro-magnetic field can directly create electron-hole pairs in the metal without need for any additional momentum-matching mechanisms. The same phenomenon can also be described by the "nonlocality", i.e., spatial dispersion of the metal dielectric constant [27,28]. While simple phenomenological picture of surface collision damping relates it to the size of the metal nanoparticles, i.e., to the electron confinement, quantum description [29] shows that the damping is determined by the photon confinement, i.e. the spatial extent of the electric field inside the metal nanoparticle. For the fundamental (dipole) SPP mode, the spatial extent of the field inside the nanoparticle is equal to its size, so two approaches give exactly the same result. In the higher order modes, however, the field is confined near the surface, and the spatial extent of the mode of the $l$-th order decreases, roughly as $(l+1)^{-1}$ shown in Section 2 below. It is then reasonably to expect the surface damping rate of the $l$-th order mode to increase as $\gamma_s^{(l)} \sim v_F(l+1)/(2a)$ where $a$ is the radius of the metal nanosphere, causing the higher order modes to broaden in the spectral domain. That in turn strongly affects the rate of coupling of energy from the emitters placed near the surface and have impact on quenching. In this work we evaluate this impact and come to conclusion that surface collisions considerably increase the quenching and reduce the amount of attainable enhancement of the luminescence.

## 2. Theory of confinement-dependent damping

Consider the spherical metal nanoparticle of radius $a$ embedded in a medium with dielectric constant $\varepsilon_d > 0$ shown in Fig.1(a). Specifically, we shall treat the nano-structure in which either Au or Ag nanospheres are embedded in the wide bandgap GaN. Each nanoparticle supports a large number of SPP modes $(l,m)$ with resonant frequencies $\omega_l = \omega_p/[1+(1+1/l)\varepsilon_d]^{1/2}$, ranging from the $\omega_1 = \omega_p/(1+2\varepsilon_d)^{1/2}$ for the lowest order radiating dipole mode to $\omega_{sp} = \omega_p/(1+\varepsilon_d)^{1/2}$ for the highest order modes that behave as surface plasmons propagating along the nanoparticle surface.

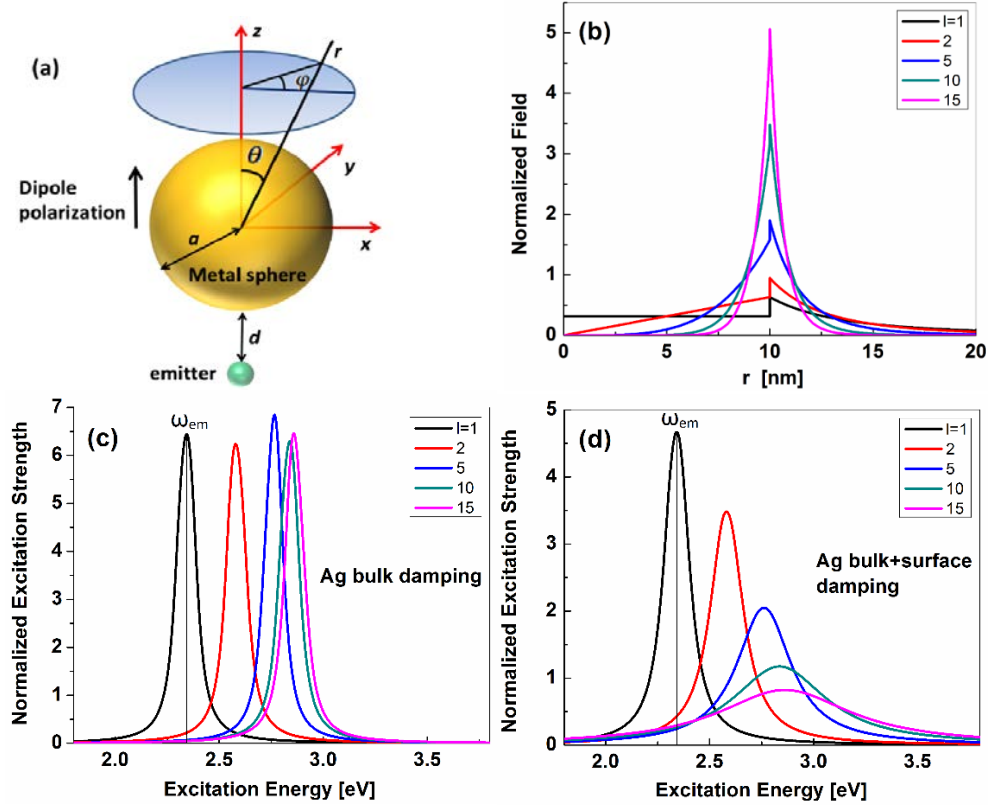

Fig. 1. (a) Geometry of luminesce enhancement by a spherical metal nanoparticle of radius $a$ separated from the emitter by distance $d$. (b) Radial electric field of SPP modes of different order $l = 1, 2, 5, 10, 15$ vs. distance from the center of nanoparticle of radius $a = 10$ nm. (c) Excitation spectra of those SPP modes for Ag nanoparticle of radius $a = 10$ nm without taking surface collision damping $\gamma_s$ into account. (d) Same as (c) but with surface collision damping taken into account.

If the polar axis z is associated with the polarization of the emitter as in Fig.1(a), only modes with $m = 0$ have non-vanishing field along this axis,

$$E_l = \begin{cases} E_{\max,l}\left(\dfrac{r}{a}\right)^{l-1}\left[-\dfrac{l}{l+1}P_l(\cos\theta)\hat{r} - \dfrac{1}{\sin\theta}[P_{l+1}(\cos\theta) - \cos\theta P_l(\cos\theta)]\hat{\theta}\right] & r < a \\ E_{\max,l}\left(\dfrac{a}{r}\right)^{l-1}\left[P_l(\cos\theta)\hat{r} - \dfrac{1}{\sin\theta}[P_{l+1}(\cos\theta) - \cos\theta P_l(\cos\theta)]\hat{\theta}\right] & r > a \end{cases} \quad (1)$$

where $\hat{r}$ and $\hat{\theta}$ are unit vectors in polar coordinates, and $P_l(\cos\theta)$ is a Legendre polynomial. The radial dependence of the electric field for different modes is shown in Fig.1(b) along z axis ($\theta = 0$) where the energy of each mode has been normalized to $\hbar\omega_l$. Since the higher order mode occupies progressively smaller volume $V_{eff,l} \approx 4\pi a^3/(l+1)^2\varepsilon_d$ [20] the amplitudes of higher order modes increase with $l$ and if an emitter is placed close enough to the metal surface to interact with these dark high-$l$ modes the energy will be very efficiently transferred into these modes and subsequently dissipated. But in order for that to happen, just having spatial overlap is not enough as the spectral overlap is also required, as shown in Fig1(c) for Ag nanosphere where each mode is represented by Lorentzian with full width at half maximum (FWHM) equal to damping $\gamma_l$ which for $l=1$ mode is equal to $\gamma_{l=1} = \gamma_{rad} + \gamma_{met}$ where $\gamma_{rad}$ is the radiative decay rate of the nanoparticle proportional to the nanoparticle volume [20], and for all other modes $\gamma_{l>1} = \gamma_{met}$. All the modes have the same bulk damping rate $\gamma_{met}$, but the surface collisions are far more prevalent in the higher order modes. One can make a very rough estimate of these rates by evaluating the mean penetration depth of the radial field into the metal as

$$w_{eff,l} = \langle (a-r)E_l^2(r)r^2 \rangle_r = \frac{a}{l+1} \tag{2}$$

which should result in surface collision damping rate of $\gamma_{s,l} \approx A(l+1)v_F/(2a)$ where according to [22] $A$ is constant of the order of unity.

To obtain a more precise value of the dependence of $\gamma_s$ on the order of the mode $l$ we consider the Lindhard formula for the longitudinal dielectric constant of the metal [29]

$$\varepsilon(\omega,k) = \varepsilon_b + \frac{3\omega_p^2}{k^2 v_F^2}\left[1 - \frac{\omega}{2kv_F}\ln\frac{\omega+kv_F}{\omega-kv_F}\right] \tag{3}$$

where $\omega_p$ is a plasma frequency and $v_F$ is the Fermi velocity of the metal. For the small wavevectors $|k| < \omega/v_F$ the dielectric constant is real, but once $|k| > \omega/v_F$, the logarithm in (3) becomes imaginary number $-i\pi$, and the dielectric constant becomes complex, acquiring imaginary part $\varepsilon_i(\omega,k) = \varepsilon_i(\omega,k) = 3\pi\omega_p^2\omega/2k^3 v_F^3$. Thus in the spatial Fourier transform of the $l$-th mode $\mathbf{E}_l(\mathbf{k})$ taken inside the metal the longitudinal spectral components with $|k| > \omega/v_F$ will get absorbed, as shown in Fig.2(a), and the surface collision contribution to the imaginary part of the effective dielectric constant of this mode will become

$$\varepsilon_{i,s}^{(l)} = \frac{3\pi\omega_p^2\omega}{2v_F^3}\left[f_x + f_y + f_z\right] \tag{4}$$

where

$$f_x = \iiint_{k_x > \omega/v_F} \frac{|E_{l,x}(k_x,k_y,k_z)|^2}{k_x^3} dk_x dk_y dk_z \tag{5}$$

with similar expressions for $f_y$ and $f_z$, and the power spectrum has been normalized as $\iiint |\mathbf{E}_l(\mathbf{k})|^2 d^3k = 1$. Using Drude formula, the surface contribution to imaginary part of dielectric constant can be related to the effective surface scattering rate as $\varepsilon_{i,s}^{(l)} = \omega_p^2 \gamma_s^{(l)}/\omega^3$. Normalizing the wavevector to the diameter of the nanoparticle, $q = 2ak_x$ and introducing $p = 2a\omega/v_F$ we obtain from Eq.(4) $\gamma_s^{(l)} = (v_F/2a)A^{(l)}(p)$, where $A^{(l)}(p) = A_x^{(l)}(p) + A_y^{(l)}(p) + A_z^{(l)}(p)$,

$$A_x^{(l)}(p) = \frac{3}{2}\pi p^4 \int_p^\infty \frac{|E_x^{(l)}(q_x)|^2}{q_x^3} dq_x \tag{6}$$

And the dimensionless $|E_x^{(l)}(q_x)|^2 = \iint |E_{l,x}(q_x,q_y,q_z)|^2 dq_y dq_z$ is normalized as $\int_0^\infty |E_x^{(l)}(q_x)|^2 dq_x = 1$. Fermi velocity in gold or silver is $v_F \approx 1.4 \times 10^6 m/s$. For the frequencies in the visible range (say 500 nm) $v_F/\omega \approx 0.37\ nm$ hence for $a > 2nm$ we have $p \gg 1$. If

longitudinal field $E_x^{(l)}(x)$ is confined on the scale $\delta x < a$ then its power spectrum $\left|E_x^{(l)}(q_x)\right|^2$ is expected to decay at large values of $q_x \gg 1$ as $K_x^{(l)} q_x^{-2}$ and integration will yield $A_x^{(l)} = 3\pi K_x^{(l)}/8$ that does not depend on $p$ and $\gamma_s^{(l)} = (v_F/2a)A^{(l)}$. To check this proposition we performed numerical calculations of $A_{x,y,z}^{(l)}(p)$ for a wide range of $p$ (corresponding to the range of $a$ from 2 to 100nm) and found that indeed these coefficients are constants and hence their sum $A^{(l)}$ is just a number, plotted in Fig.2 b. It is clear that the coefficient $A^{(l)}$ can be very well approximated as $l$ and the collision scattering damping as $\gamma_s^{(l)} \approx l v_F/2a$ which is only slightly different from the phenomenological estimate made from Eq. (2). The result of this surface damping adding on to the bulk damping rate leads to increase of Lorentzian FWHM of various SPP modes as shown in Fig.1(d) with higher order ones being progressively broadened.

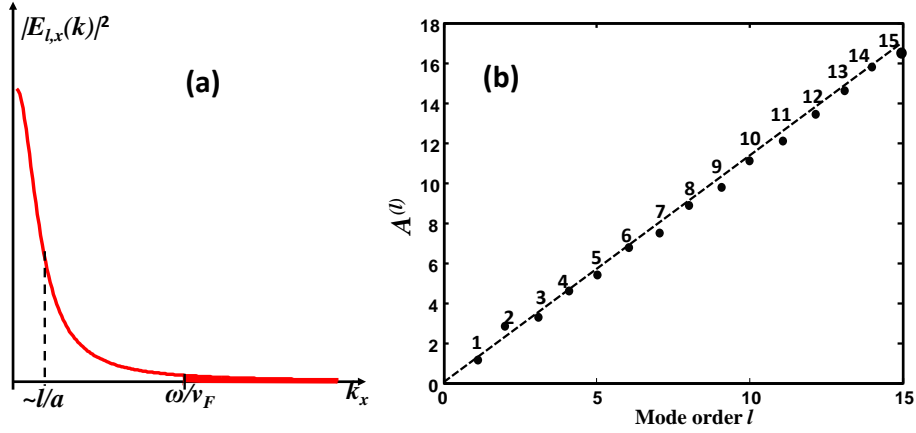

Fig. 2. (a) Power spectrum of the $l$-th order SPP mode. The fraction of energy with wave vector larger than $\omega/v_F$ get absorbed by the metal (Landau damped). (b) Calculated values of damping coefficient $A^{(l)}$ for SPP modes from $l=1$ to $l=15$. Linear fit $A^{(l)} = l$ is shown by the dashed line.

## 3. Impact on the Quenching Coefficient

In comparison with Fig.1(c), the excitation spectra of different SPP modes of Ag nanosphere with surface collision broadening taken into account as shown in Fig. 1(d) shows that higher order modes get broadened considerably. For example, spectrum of $l=10$ mode broadens by a factor of 5 and its overlap with the emission line of the emitter, centered at frequency $\omega_{em}$ increases accordingly. The coupling of the emitter's energy into the $l$-th order SPP mode is characterized by the Purcell factor [20],

$$F_P^{(l)} = \frac{3\pi\varepsilon_d (l+1)^2 \omega_{em} L_l(\omega_{em})}{4}\left(\frac{c}{\varepsilon_d^{1/2}\omega_{em} a}\right)^3 \left(\frac{a}{a+d}\right)^{2l+4}, \qquad (7)$$

where Lorentzian line shape of the mode is $L_l(\omega_{em}) = (\gamma_l/2\pi)/\left[(\omega_{em}-\omega_l)^2 + \gamma_l^2/4\right]$. Since the emission line is typically resonant with the dipole mode, $\omega_{em} = \omega_1$, it is clear that for as long as $\omega_l - \omega_1 > \gamma_l/2$, i.e. practically for all the dark modes with $l \geq 2$, the increase in damping causes increase in Purcell factor $F_P^{(l)}$. For the $l=1$ bright dipole mode on the other hand, increase in dumping causes reduction of $F_P^{(1)}$. The changes caused by increased surface collision damping is well described by the "quenching" ratio [18] $f_q = \sum_{l=2}^{\infty} F_P^{(l)} / F_P^{(1)}$, i.e., the ratio of summation of all Purcell factors associated with all the dark modes ($l \geq 2$) to the Purcell factor for the "bright" dipole mode ($l = 1$). The quenching factors for gold and silver nanospheres with radius $a = 5$ nm are plotted versus the separation distance of the emitter $d$ in Figs.3(a) and (b) respectively. The "Bulk" curves in each figure are describing the quenching under assumption that only the "bulk" non-radiative damping $\gamma_{met}$ is contributing. The curves "Surface" include only the surface contribution to damping $\gamma_s^{(l)}$ and the curves "Bulk + Surface" correspond to the realistic description of damping that includes both bulk and surface contributions.

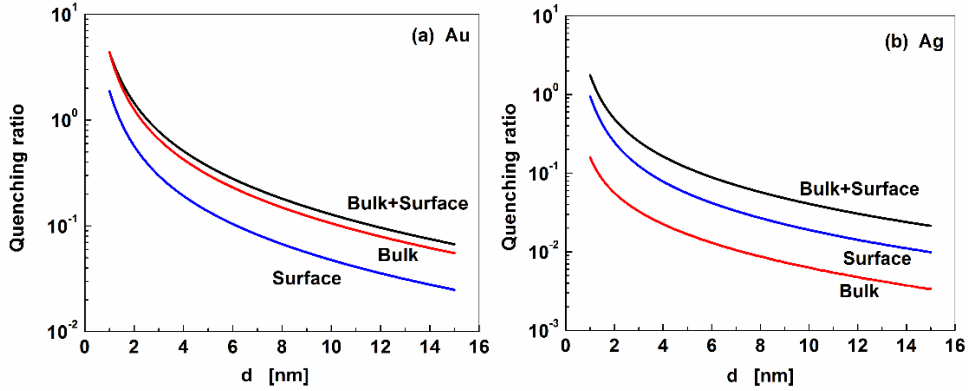

Fig. 3 Quenching ratios as a function of distance $d$ between the emitter and metal nanosphere of radius $a = 5$ nm with SPP mode damping rate taking into account bulk nonradiative or surface collision damping or both combined for (a) Au and (b) Ag.

Other than the obvious quenching effect being stronger for smaller separation between the metal nanoparticle and the emitter, quenching increases when surface collision damping is taken into account. While the quenching ratio for Ag is generally smaller than that of Au because of its smaller ohmic losses, the impact of surface damping is more significant for Ag given the fact that both nanoparticles of equal size have similar surface collision damping rate $\gamma_s^{(l)} \approx l v_F / (2a)$, but this rate represents a high fraction of the total damping rate $\gamma_l$ for Ag. Another interesting observation is that even if we had ways to magically eliminate the "bulk" ohmic losses in the metal, we would still have been stuck with surface damping and quenching would still take place because this surface effect is inherently associated with the field localization of the various orders of the SPP modes.

### 4. Optimization of Luminescence Enhancement

Once the quenching factor had been determined one can evaluate the luminescence enhancement factor for an emitter with an original radiative efficiency $\eta_{rad}$ placed in the vicinity of meal nanoparticle [18]

$$F = \frac{1 + F_P^{(1)} \eta_{pr}}{1 + F_P^{(1)}(1 + f_q)\eta_{rad}} \quad (8)$$

where $\eta_{pr} = \gamma_{rad}/\gamma_1$ is the out-coupling efficiency of dipole mode. All the factors ($F_P^{(1)}$, $\eta_{pr}$, and $f_q$) depend on the nanosphere radius $a$. Hence one can perform an optimization to determine the ideal size of the nanosphere that is small enough to have a small effective mode volume $V_{eff,l}$ assuring large Purcell factor of the dipole SPP mode $F_P^{(1)}$ in (7), yet is still sufficiently large to assure a decent radiative out-coupling efficiency $\eta_{pr}$ of this mode. Furthermore, according to (7) if the quenching through high order modes could be neglected, it would have been always better to have emitters placed as close as possible to the metal nanoparticle. But as shown in previous section the quenching also increases as the separation distance $d$ gets shorter. Hence the distance $d$ can also be optimized to achieve a favorable compromise of good coupling into the dipole mode and adequate suppression of the luminescence quenching by higher order modes. The dependence of the enhancement by Au nanosphere on $a$ and $d$ is shown in inset of Fig.4(a) for the emitter with $\eta_{rad} = 0.01$. The results of optimization of luminescence enhancement for Au and Ag nanospheres are shown in Fig.4(a) and (b) respectively. The optimization is performed for a wide range of emitter radiative efficiencies $\eta_{rad}$. The optimal sphere radius $a_{opt}$ and separation $d_{opt}$ at which the maximum attainable enhancement is obtained for each $\eta_{rad}$ are shown in Fig. 4(c-f). The three graphs in each plot correspond to the same three cases of damping as in Fig.3.

As expected [16], the maximum attainable enhancement deteriorates quite rapidly with the increase of $\eta_{rad}$. With the inclusion of surface collision damping, the attainable enhancement is reduced for both Au and Ag nanoparticles, but the impact is much stronger for Ag for the same reason as explained in section 3. Also, the optimized values of radius $a_{opt}$ and separation distances $d_{opt}$ also increase once surface collision damping is taken into account. The increase in $a_{opt}$ is easy to understand – the surface collision damping is inversely proportional to the radius. The modest increase in $d_{opt}$ obviously reduces coupling into the broadened higher order modes, while keeping coupling into the dipole mode reasonably high.

As one can see that for noble metal spheres the changes imposed by the surface collision damping are significant. For Ag the quenching ratio [Fig 3(b)] is increased by an order of magnitude and the maximum enhancement [Fig 4(b)] is reduced by a factor of 3.5. The reason for this is that even when Ag sphere is embedded in high index dielectric the dipole SPP resonance is near 2.3eV where the bulk damping rate in Ag $\gamma_{met} \approx 1.2 \times 10^{14} s^{-1}$ [30] since that energy is close to the onset of interband transitions. Since the surface collision rate is roughly $\gamma_s^{(l)} \approx l v_F / 2a$, in the 10 nm radius nanosphere this rate is comparable to the bulk rate for the quadrupole mode $l = 2$. If, however, one uses nanoshells [31] or dimers [32] the SPP resonances can be moved to the red part of the spectrum, say 700nm where $\gamma_{met} \approx 3.2 \times 10^{13} s^{-1}$ and obviously the influence of surface collision damping will grow four-fold leading to much larger changes in quenching ratios.

Once again, it should be noted that even in the perfect world where one could eliminate all "bulk" ohmic loss of any metal, the enhancement would never be "perfect", as will be limited by the surface damping as shown in Fig. 4(a) and (b) for both Au and Ag nanoparticles, respectively.

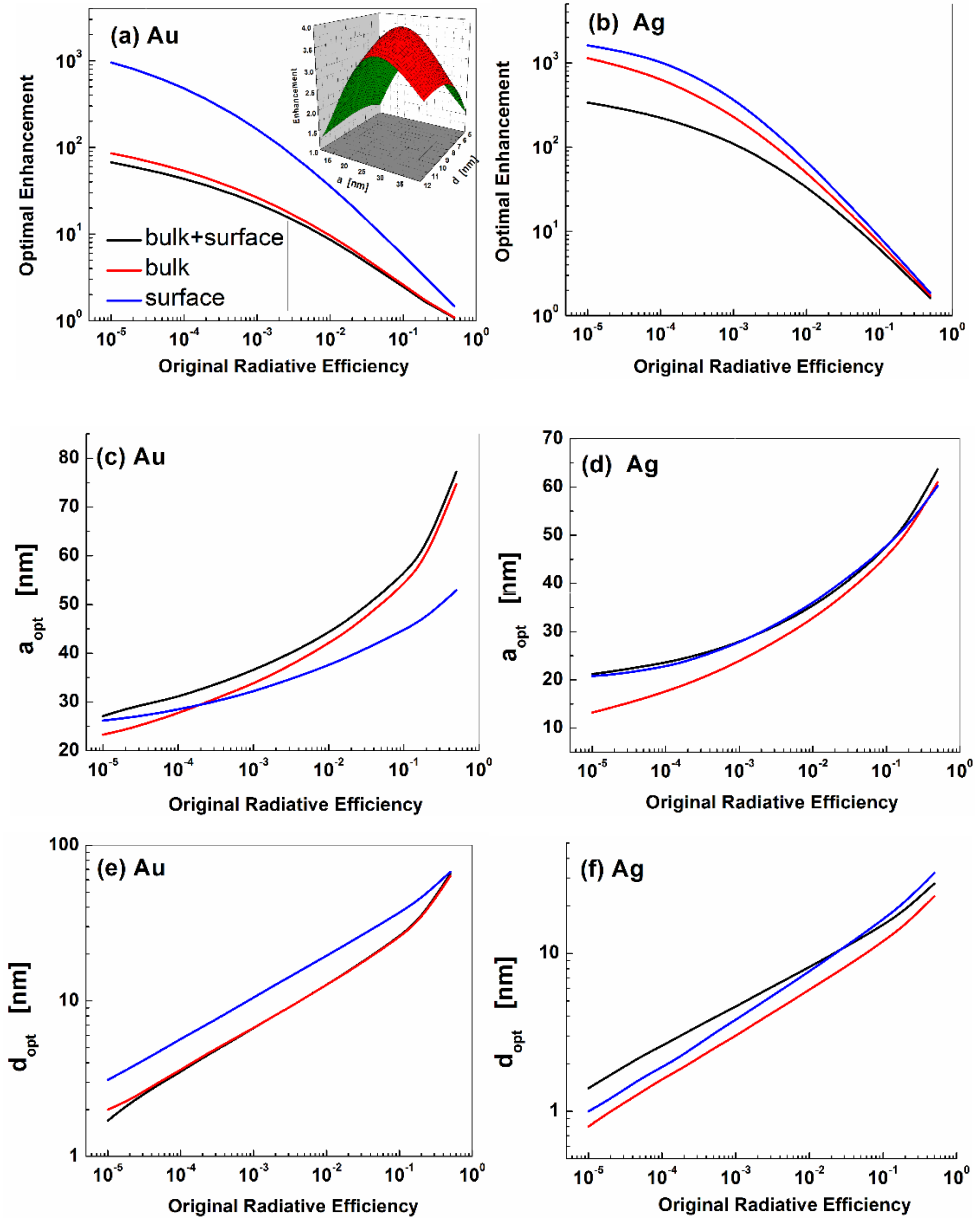

Fig. 4 Optimized enhancement, nano-sphere radius $a_{opt}$, and molecule-sphere separation $d_{opt}$ over a range of original radiative efficiency for Au (a), (c), (e) and Ag (b), (d), (f), respectively.

## 5. Conclusion

In conclusion, we have shown that high order SPP modes in metal nanoparticles are subject to much stronger damping than the lowest order dipole mode. This additional damping, roughly proportional to the order of the mode is caused by the confinement of the higher order mode in the vicinity of the surface. Increased damping causes spectral broadening of the higher order SPP resonances and increases the probability of their excitation by the luminescing object placed near the nanoparticle. Since the higher order modes are dark the energy gets trapped and eventually dissipated in them and the luminescence experiences additional quenching. As result, maximum enhancement of luminescence attainable in the vicinity of small metal nanoparticles gets reduced by a factor of 3 or 4 in the blue green range and even larger factors in the red part of spectrum.

**Acknowledgments**

JK acknowledges fruitful discussions with Dr. P. Noir. GS acknowledges support from AFOSR (FA9550-14-1-0196 Dr. Gernot Pomrenke, Program Manager).